\RequirePackage{fix-cm}
\documentclass[smallextended]{svjour3}       
\smartqed  
\usepackage{graphicx}
\usepackage{epstopdf}
\usepackage{cite}
\usepackage{amsmath,amssymb,amsfonts}
\usepackage{algorithmic}
\usepackage{graphicx}
\usepackage{textcomp}
\usepackage{caption}
\usepackage{subfigure}
\usepackage{multirow}
\usepackage{multicol}
\usepackage{booktabs}
\usepackage{color}
\begin{document}

\title{Quantum Forgery Attacks on COPA, AES-COPA and Marble Authenticated Encryption Algorithms 
}
\subtitle{}


\author{Yinsong Xu         \and
        Wenjie Liu* \and Wenbin Yu 
}


\institute{Y. Xu \at
              School of Computer and Software, Nanjing University of Information Science and Technology, Nanjing 210044, P. R. China \\
              \email{mugongxys@foxmail.com}             \\
           \and
           W. Liu \at
           *Corresponding Author\\
              Engineering Research Center of Digital Forensics, Ministry of Education, Nanjing 210044, P. R. China\\
              School of Computer and Software, Nanjing University of Information Science and Technology, Nanjing 210044, P. R. China\\
              \email{wenjiel@163.com}             \\
              \and
              W. Yu\at
              Engineering Research Center of Digital Forensics, Ministry of Education, Nanjing 210044, P. R. China\\
              School of Computer and Software, Nanjing University of Information Science and Technology, Nanjing 210044, P. R. China\\
              \email{ywb1518@126.com}             \\
}

\date{Received: date / Accepted: date}

\maketitle

\begin{abstract}
Since the classic forgery attacks on COPA, AES-COPA and Marble authenticated encryption algorithms need to query about ${2^{n/2}}$ times and their success probability are not high. To solve this problem, the corresponding quantum forgery attacks on COPA, AES-COPA and Marble authenticated encryption algorithms are presented. In the quantum forgery attacks on COPA and AES-COPA, we use Simon's algorithm to find the period of the tag generation function in COPA and AES-COPA by querying in superposition, and then generate a forged tag for a new message. While in the quantum forgery attack on Marble, Simon's algorithm is used to recover the secret parameter $L$, and the forged tag can be computed with $L$. Compared with classic forgery attacks on COPA, AES-COPA and Marble, our attack can reduce the number of queries from $O({2^{n/2}})$ to $O(n)$ and improve success probability close to 100\%.

\end{abstract}

\section{Introduction}
\label{sec:1}
In symmetric cryptography, an authenticated encryption algorithm is an algorithm that transforms an arbitrary-length data stream, called a message or plaintext, into another data stream of the same length, called a ciphertext, and generates an authentication tag for the message at the same time, under the control of a secret key \cite{Lu16}. The purpose of authenticated encryption algorithms is to provide data privacy and integrity. In 2013, the CAESAR competition \cite{CAESAR} was launched to provide good authenticated encryption schemes as better alternatives to current options such as AES-GCM \cite{Boer79}, which have received 57 submissions in first round.

The COPA authenticated encryption algorithm \cite{Andreeva13} is designed by Andreeva \textit{et al.}, which combines OCB's offsets with an internal dependency chain in order to achieve some security in the case of nonce repetition. Since its birth is earlier than the CAESAR competition, it did not participate in the competition. But, its instantiation with the AES \cite{AES} block cipher under 128 key bits (called AES-COPA \cite{AES-COPAv1,AES-COPAv2}) has been a CAESAR candidate. The Marble authenticated encryption algorithm (v1.0/1.1/1.2) \cite{Marble-v1.0,Marble-v1.1,Marble-v1.2} is also a CAESAR submission by Jian Guo inspired by COPA, which uses two internal chains to prevent birthday attacks on the internal chain and uses reduced-round AES as building blocks. So far, only AES-COPA is a CAESAR candidate.

Although the COPA designers proved that it has a birthday-bound security on integrity (which is mainly associated with existential forgery) under the assumption that the underlying block cipher is a strong pseudorandom permutation. And the Marble designer claimed that Marble achieved a full security beyond the birthday-bound due to the choice of $TRANS$ function (see Table \ref{Tab:1}). However, Nandi \cite{Nandi15} presented an existential forgery attack on the case of COPA that processes fractional messages, which produce the correct ciphertext and tag for an unspecified message whose ciphertext and tag are not given. Later, Lu \cite{Lu17} also proposed an almost universal forgery attack on the COPA and Marble, which produce the correct ciphertext and tag for any specified message whose ciphertext and tag are not given. Note that almost universal forgery attack allows the forger to make a minimal change from the given message $M$ to a modified one $M'$ by replacing some of its blocks before producing its tag \cite{Dunkelman15}. Both of Nandi's and Lu's forgery attack on the COPA indicate that the probability of a forgery is much larger than $2/{2^n}$ when the number of queries $q$ is close to birthday-bound ${2^{n/2}}$ and even smaller than ${2^{n/2}}$.

Moreover, Lu's forgery attack on the marble also indicates that the Marble (v1.0/1.1/1.2) are incorrectly far overestimated in the sense of full security. The probability of a forgery would be 32\% when the number of queries $q$ is close to birthday-bound ${2^{n/2}}$. Due to that Lu's forgery attack on the Marble do not use associated data, Fuhr \textit{et al.}'s forgery attack \cite{Fuhr15} make up for this consideration. Finally, for AES-COPA with nonce, Lu's forgery attack requires slightly less than ${2^{63}}$ encryption queries and its success probability is about 6\%.

On the other hand, in the quantum world, many quantum algorithms are constantly being used in cryptanalysis \cite{Shor94,Kuwakado10,Kuwakado12}, machine learning \cite{Liu18,Biamonte17,Liu20}, blockchain \cite{Gao20,Banerjee20} and so on. Since Shor's algorithm \cite{Shor94} was proposed, it has been announced that quantum computers would be a severe threat for public key cryptography. More and more researchers have began to use quantum algorithms to break symmetric cryptosystems, such as Simon's algorithm \cite{Simon97,Kuwakado10,Kuwakado12,Kaplan16,Shi18}, Grover algorithm \cite{Grover97,Leander17}, Bernstein-Vazirani algorithm \cite{Bernstein97,Xie19} and so on. In addition, they also have proposed some new quantum algorithms \cite{Chailloux17,Hosoyamada17,Hosoyamada19}, and even extended classical cryptanalysis methods to the quantum field \cite{Bonnetain19,Hosoyamada18}. Among them, Simon's algorithm was first used to break the 3-round Feistel construction \cite{Kuwakado10} and then to prove that the Even-Mansour construction \cite{Kuwakado12} is insecure with superposition queries. Inspired from them, Kaplan \textit{et al.} \cite{Kaplan16} show that several classical attacks based on finding collisions can be dramatically sped up using Simon's algorithm. And Shi \textit{et al.} \cite{Shi18} also use analogous way to implement collision attacks on authenticated encryption AEZ from CAESAR competition. To improve the efficiency of classic forgery attack on COPA, AES-COPA and Marble, we present quantum forgery attacks based on Simon's algorithm. Note that, the attack of Ref. \cite{Kuwakado10,Kuwakado12,Kaplan16,Shi18,Leander17,Bernstein97,Xie19,Chailloux17,Hosoyamada17,Hosoyamada19,Bonnetain19,Hosoyamada18} and our forgery attack all belong to the Q2 model proposed by Kaplan \cite{Kaplan14}. In Q2 model, the adversary is allowed to perform quantum superposition queries to a remote quantum cryptographic oracle \cite{Chailloux17}. The opposite of Q2 model is Q1 model, i.e., the adversary can query a quantum random oracle with arbitrary superpositions of the inputs, but is only able to make classical
queries to a classical encryption oracle. Therefore, this model is not considered in this article.

\textbf{Our contributions.} In this paper, we use Simon's algorithm to find the period of the function of tag generation in COPA firstly. When we get the period, we can compute the forged message for the same tag. Since the length of the associated data block and the message block will affect the period value, there will be a small difference in the process of applying Simon's algorithm (see Sect. \ref{sec:3}). Secondly, the encryption of AES-COPA is analogous to COPA, an additional (public) input parameter called nonce is appended to associated data (if any), and then the resulting value is treated as associated data in COPA. Therefore, our quantum forgery attack on AES-COPA is similar on COPA. Moreover, due to the tag generation has 3-round encryption with different keys and the function $TRANS$, it is difficult to find the period of tag by Simon's algorithm. But, we use Simon's algorithm to find the period of internal state ${S_1}$, and then obtain the secret parameter $L$ through the got period. Finally, we can use the secret parameter $L$ to compute forged tag and message pairs. In summary, the query times and the success probability of our quantum forgery attack are mainly reflected in the number of executing Simon's algorithm and the success probability of finding period. That is, our quantum forgery attack only needs $O(n)$ queries with high success probability (the probability is close to 1).

This paper is organized as follows. Sect. \ref{sec:2} provides a brief description of the COPA, AES-COPA, Marble authenticated encryption algorithms and Simon's algorithm. And our quantum forgery attacks on COPA, AES-COPA and Marble authenticated encryption algorithms are shown in Sects. \ref{sec:3}, \ref{sec:4} and \ref{sec:5}, respectively. Then, the comparison with other forgery attacks on COPA, AES-COPA and Marble is analyzed in Sect. \ref{sec:6}, followed by a discussion and conclusion in Sect. \ref{sec:7}.

\section{Preliminaries}
\label{sec:2}
In this section, we would briefly describe the COPA, AES-COPA, Marble authenticated encryption algorithms and Simon's algorithm.
\subsection{COPA, AES-COPA and Marble Authenticated Encryption Algorithms}
\label{sec:2_1}
\begin{figure*}
\centering
  \includegraphics[width=4.5in]{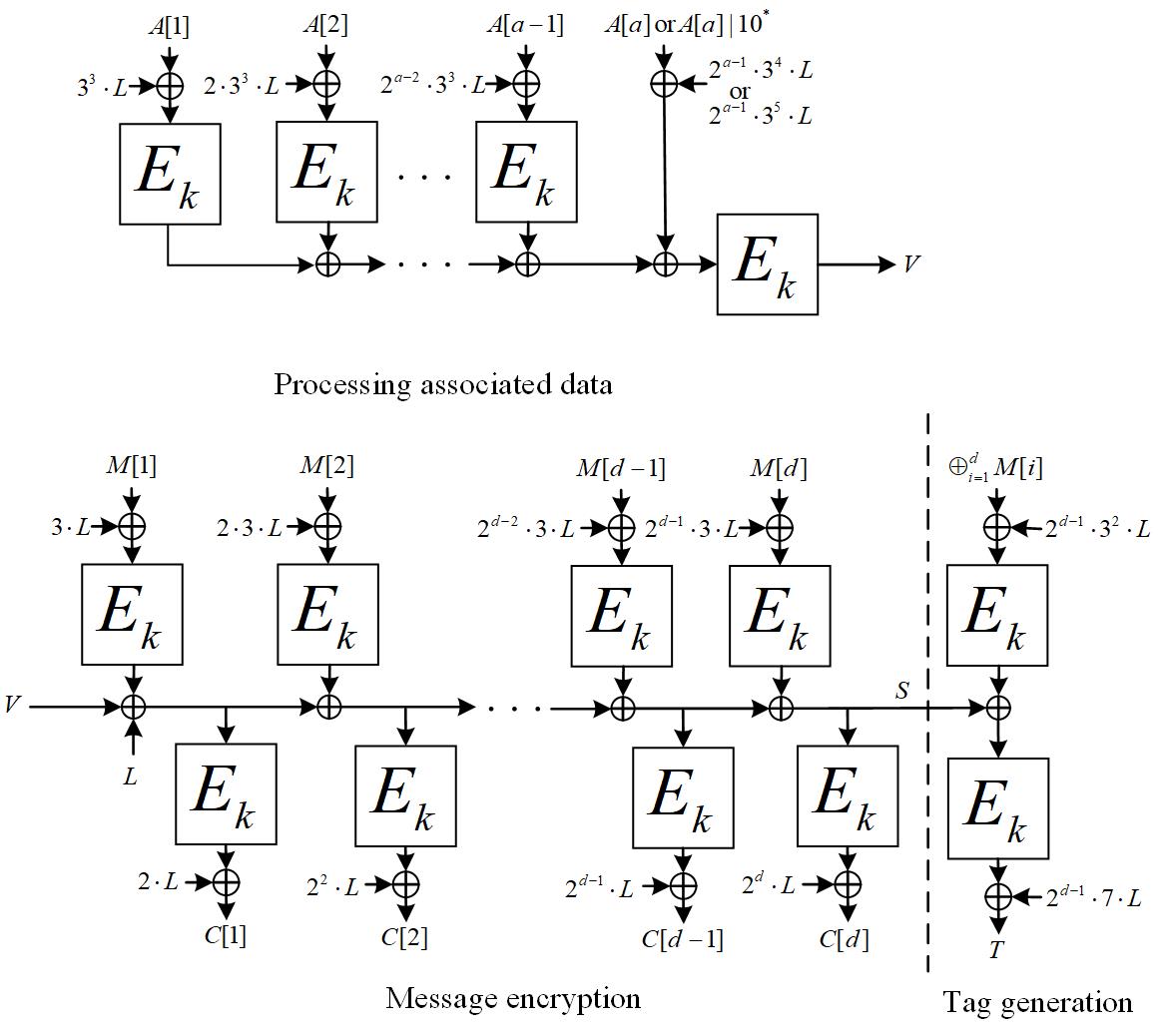}
\caption{The process of processing associated data, message encryption and tag generation in COPA.}
\label{fig:1}       
\end{figure*}

For the sake of clarity, we list some explanations for variables and notations in Table \ref{Tab:1}, which are frequently used in COPA and Marble authenticated encryption algorithms. The COPA authenticated encryption algorithm \cite{Andreeva13} was published in 2013. Its internal state, key and tag have the same length. Therefore, in order to facilitate following analysis, we default the length of them to 128 bits. To generate ciphertexts and tag, it has three phases: processing associated data, message encryption, and tag generation, which is shown in Fig. \ref{fig:1}. During the process of processing associated data, if there is no associated data, then we set $V\mathop  = \limits^{def} 0$. Besides, if the last block $A[a]$ or $M[d]$ are not a multiple of $n$ bits, they need to be padded by a one and as many zeroes as necessary to obtain a multiple of the block size $n$, i.e., $A[a]{10^*}$ and $M[d]{10^*}$. Finally, decryption is the inverse of encryption, and tag verification is identical to tag generation. Please refer to \cite{Andreeva13} for the specification of COPA.

\begin{table}[ht]
  \caption{Variables and notations}
  \label{Tab:1}
  \centering
  \begin{tabular}{p{3cm}p{8cm}}
  \toprule
  \textbf{Variables and notations} &\textbf{Explanations} \\
  \midrule
  $A[1]||A[2]|| \cdots ||A[a]$, $M[1]||M[2]|| \cdots ||M[d]$        &  $A[1]||A[2]|| \cdots ||A[a]$ and $M[1]||M[2]|| \cdots ||M[d]$ are represented as associated data of $a$ $n$-bit blocks and messages of $d$ $n$-bit blocks, respectively, where "$||$" is bit connection and $n$ generally defaults to 128.	 \\
  $C[1],C[2], \cdots ,C[d]$; $T$ & $C[1],C[2], \cdots ,C[d]$ and $T$ are the ciphertext and the tag for $M[1]$$||$$M[2]$$||$$ \cdots $$||$$M[d]$, repestively.\\
  $\left| A \right|$& $\left| A \right|$ represents the number of bits in $A$.\\
  $S$, ${S_1}$, ${S_2}$ & $S$, ${S_1}$ and ${S_2}$ are $n$-bit ($n=128$) internal states.\\
  ${E_k}()$, $L$ & ${E_k}()$ is an $n$-bit block cipher, i.e., $E:k \times {\{ 0,1\} ^n} \to {\{ 0,1\} ^n}$, where the key $k$ generally consists of 128 bit. And $L\mathop  = \limits^{def} {E_k}(0)$ in COPA.\\
  $ + $, $ \oplus $ & "$ + $" or "$ \oplus $" are bitwise logical exclusive (XOR) operation.\\
  $ \cdot $ & "$ \cdot $" represents polynomial multiplication modulo the polynomial ${x^{128}} + {x^7} + {x^2} + x + 1$ in GF(${2^{128}}$). We can abbreviate $A \cdot B$ as $AB$.\\
  ${E_1}$, ${E_2}$, ${E_3}$ & Each of the operations ${E_1}$, ${E_2}$ and ${E_3}$ is a 4-round reduced version of the AES [29] block cipher, with four fixed round subkeys chosen from the eleven round subkeys of the AES with 128 key bits.\\
  \includegraphics [width=0.1in]{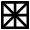} & \includegraphics [width=0.1in]{Trans}  (see Fig. 2) represents a function $TRANS(x,y) = (x + y,3x + y)$, where $x$ and $y$ are 128-bit inputs.\\
  $Cons{t_0}$, $Cons{t_1}$, $Cons{t_2}$ & $Cons{t_0}$, $Cons{t_1}$ and $Cons{t_2}$ are three 128-bit constants.\\
  $\tau $ & $\tau $ is 128-bit secret parameters.\\
  \bottomrule
  \end{tabular}
\end{table}

AES-COPA authenticated encryption algorithm is an extended version of COPA, with some differences. First, a public message number $N$ called nonce is appended to associated data, like $A[1]||A[2]|| \cdots ||A[a]||N$, and as part of associated data. Besides, AES-COPA can accept "fractional" messages $M$, i.e., the length $\left| M \right|$ is not necessarily a positive multiple of the block size $n$. And AES-COPA has two versions v.1 \cite{AES-COPAv1} and v.2 \cite{AES-COPAv2}, where the process of fractional message encryption in v.1 is slightly different from v.2 (as shown in Table \ref{Tab:2}). For simplicity, we roughly introduce the encryption process of these two versions. Note that $XL{S_d}()$ is invertible in v.1.

\begin{table}[ht]
\caption{ {Two versions of AES-COPA}}
  \label{Tab:2}
  \centering
  \begin{tabular}{p{5cm}|p{5cm}}
  \toprule
  \textbf{AES-COPA v.1-ENCRYPT:} & \textbf{AES-COPA v.2-ENCRYPT:}\\
  \midrule
  \textbf{if} $d  \geqslant 2$ and $1 \leqslant |M[d]| \leqslant n - 1$ \textbf{then}&  $M = M[1]||M[2]|| \cdots ||M'[d]$\\
  \quad  $V$ $ \leftarrow $ Processing associated data $A||N$ (see Fig. \ref{fig:1}) &  $M[d]$ $ \leftarrow $ $M'[d]$ \textbf{if} $\left| {M'[d]} \right| = n$ \textbf{else} $M'[d]|{10^*}$\\
  \quad   ($C',S'$) $ \leftarrow $ Message encryption ($V,M[1]||M[2]|| \cdots ||M[d - 1]$) (see Fig. \ref{fig:1}) & $P$ $ \leftarrow $  0 \textbf{if} $\left| {M'[d]} \right| = n$ \textbf{else}  1\\
  \quad $\Sigma '$ $ \leftarrow $ $M[1] \oplus M[2] \oplus  \cdots  \oplus M[d - 1]$ & $V$ $ \leftarrow $ Processing associated data $A||N$ (see Fig. \ref{fig:1}) \\
  \quad $T'$ $ \leftarrow $ ${E_k}({E_k}(\Sigma ' \oplus {2^{d - 2}}{3^2}L) \oplus S') \oplus {2^{d - 2}}7L$& ($C,S$) $ \leftarrow $ Message encryption ($V,M[1]||M[2]|| \cdots ||M[d],P$) (see Fig. \ref{fig:1})\\
  \quad $C[d]T$ $ \leftarrow $ $XL{S_d}(M[d]T')$ & $\Sigma $ $ \leftarrow $ $M[1] \oplus M[2] \oplus  \cdots  \oplus M[d]$\\
  \quad $C$ $ \leftarrow $ $C'C[d]$ & $T$ $ \leftarrow $ ${E_k}({E_k}(\Sigma  \oplus {2^{d - 1}}{3^2}{7^P}L) \oplus S) \oplus {2^d}7L$\\
  \quad Output ($C$,$T$) & Output ($C$,$T$)\\
  \textbf{end if}\\
  \bottomrule
  \end{tabular}
\end{table}

The Marble authenticated encryption algorithms \cite{Marble-v1.0,Marble-v1.1,Marble-v1.2} are like COPA, which has four phases: initialization, processing associate data, message encryption, and tag generation. Fig. 2 illustrates the message encryption and tag generation phase of newest version (i.e. v1.2) of Marble. Its decryption is the inverse of encryption, and tag verification is identical to tag generation. Please refer to \cite{Marble-v1.0,Marble-v1.1,Marble-v1.2} for the specification of Marble. The main differences between the newest version of Marble and the other two versions are as follows: In the second version (i.e. v1.1 \cite{Marble-v1.1}), the mask parameter before ${E_1}$ is ${2^{a - 1}} \cdot {3^2} \cdot L$ for the last block of associated data; and in the initial version (i.e. v1.0 \cite{Marble-v1.0}), the mask parameter before ${E_1}$ is ${2^{a - 1}} \cdot {3^3} \cdot L$ for the last block of associated data if it is full, and is ${2^{a - 1}} \cdot {3^4} \cdot L$ if it is not full. Thus, the latest version of Marble is identical to the initial version when the last block of associated data is full.

%

\begin{figure*}
\centering
  \includegraphics[width=4.5in]{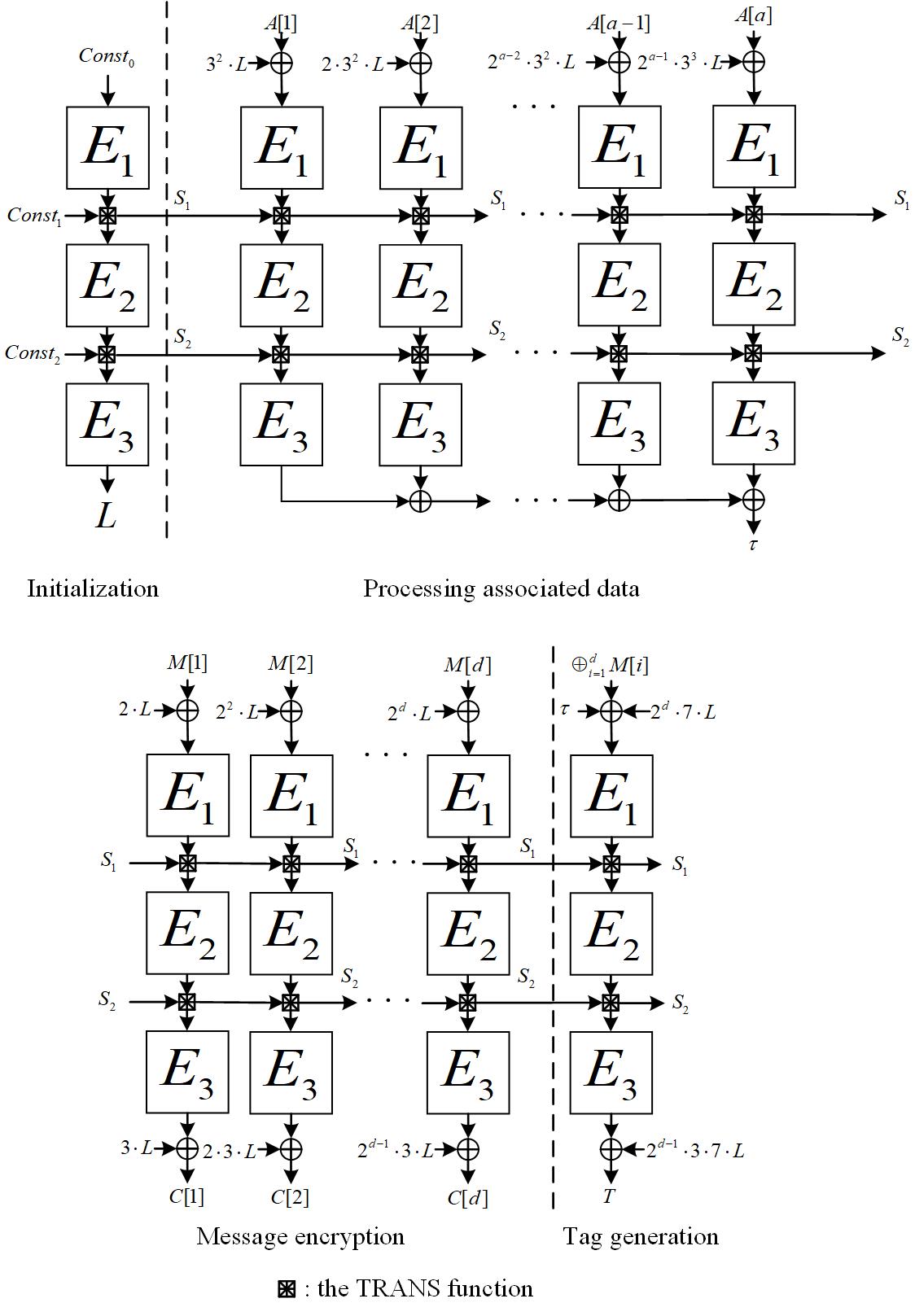}
\caption{The process of initialization, processing associated data, message encryption and tag generation in Marble.}
\label{fig:2}       
\end{figure*}

\subsection{Simon's Algorithm}
\label{sec:2_2}
Simon's algorithm was proposed by Daniel R. Simon \cite{Simon97} in 1997, which is a quantum algorithm to solve Simon's problem (also proposed by Daniel R. Simon). The definition of Simon's problem is presented as below.

\noindent\textbf{Simon's Problem}: Given a Boolean function $f:{\{ 0,1\} ^n} \to {\{ 0,1\} ^n}$ and the promise that there exists $s \in {\{ 0,1\} ^n}$ such that for any $(x,y) \in {\{ 0,1\} ^n}$, $[f(x) = f(y)] \Leftrightarrow [x \oplus y \in \{ {0^n},s\} ]$, the goal is to find $s$.

This problem can be solved classically by searching for collisions. The optimal time to solve it is therefore $\Theta ({2^{n/2}})$. On the other hand, Simon's algorithm solves this problem with quantum complexity $O(n)$. Note that to run Simon's algorithm, it is required that the function $f$ can be queried quantum-mechanically.

\begin{figure*}
\centering
  \includegraphics[width=4in]{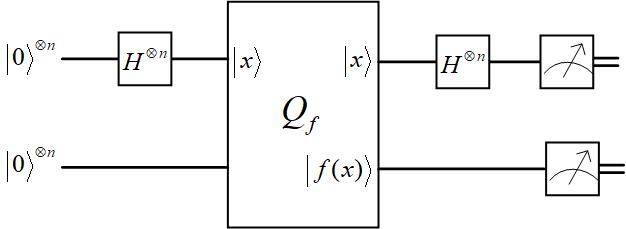}
\caption{The circuit of Simon's algorithm.}
\label{fig:3}       
\end{figure*}

In Simon's algorithm (as shown in Fig. \ref{fig:3}), we need to prepare a $2n$-qubit state ${\left| 0 \right\rangle ^{ \otimes n}}{\left| 0 \right\rangle ^{ \otimes n}}$ and apply Hadamard transform ${H^{ \otimes n}}$ to the first $n$ qubits to obtain the quantum superposition $\frac{1}{{\sqrt {{2^n}} }}\sum\limits_{x \in {{\{ 0,1\} }^n}} {\left| x \right\rangle {{\left| 0 \right\rangle }^{ \otimes n}}} $. Then, the quantum superposition would be input into the function $f$ to get the state $\frac{1}{{\sqrt {{2^n}} }}\sum\limits_{x \in {{\{ 0,1\} }^n}} {\left| x \right\rangle \left| {f(x)} \right\rangle } $. After that, we apply Hadamard transform ${H^{ \otimes n}}$ to the first $n$ qubits again to get $\frac{1}{{{2^n}}}\sum\limits_{x \in {{\{ 0,1\} }^n}} {\sum\limits_{y \in {{\{ 0,1\} }^n}} {{{( - 1)}^{x \cdot y}}\left| y \right\rangle \left| {f(x)} \right\rangle } } $. Finally, we perform measurements on all qubits, where the vector $y$ (measured from the first $n$ qubits) be orthogonal to $s$, i.e., $y \cdot s = 0$. By repeating this subroutine $O(n)$ times, one obtains $n-1$ independent vectors orthogonal to $s$ with high probability, and $s$ can be recovered using basic linear algebra.

\section{Quantum Forgery Attacks on COPA by Simon's Algorithm}
\label{sec:3}

\subsection{Attack Strategy}
\label{sec:3_1}
Due to that our forgery attack belongs to Q2 model, the adversary can be able to access the quantum cryptographic oracle and queries in superposition. To execute Simon's algorithm, the quantum cryptographic oracle would be used as quantum oracle ${Q_f}$ (also as the function $f$ in Simon's problem) in the circuit of Simon's algorithm. According to the specific situation, we select the associated data or message as the input of Simon's algorithm, and select the tag or other data (like ${S_1}$ in Marble) as the algorithm's output. By repeatedly executing Simon's algorithm $O(n)$ times, we can obtain the corresponding period value. Finally, we can get the collision of the tag with different associated data or messages by the period value. The entire attack process is called as a quantum forgery attack.



Since COPA can set the associated data to 0, we conduct quantum forgery attacks from two cases: without associated data and with associated data, which is demonstrated as below.

\subsection{Quantum Forgery Attacks on COPA without  {Associated Data}}
\label{sec:3_2}
Since there is no associated data, so $V = 0$ and $T = {E_k}({E_k}(\Sigma  \oplus {2^{d - 1}}{3^2}L) \oplus S) \oplus {2^{d - 1}}7L$, where $\Sigma  = M[1] \oplus  \cdots  \oplus M[d]$. We can see that the size of the message length $d$ affects the period value $s$ of the function $T$. Therefore, we will calculate the period $s$ of $T$ from 2 cases: $d = 1$ and $d \geqslant 2$.

\noindent \textbf{Case 1}: When $d = 1$, i.e., message $M = M[1]$. Then,
\begin{equation}\label{eq01}
T = {E_k}({E_k}(M[1]\oplus {3^2}L) \oplus S) \oplus 7L,
\end{equation}
where $S = {E_k}(M[1] \oplus 3L) \oplus V \oplus L = {E_k}(M[1] \oplus 3L) \oplus L$ and ${3^2}= 3 \cdot 3=(x + 1)(x + 1)=({x^2} + x + x + 1) mod({x^{128}} + {x^7} + {x^2} + x + 1)={x^2}+1=5$. Note that, the multiplication and addition operations in this paper are the addition and multiplication operations in GF(${2^{128}}$). So,

\begin{equation}\label{eq02}
T = {E_k}({E_k}(M[1] \oplus 5L) \oplus {E_k}(M[1] \oplus 3L) \oplus L) \oplus 7L.
\end{equation}

Firstly, we define the following function:
\begin{equation}\label{eq03}
\begin{split}
f:{\{ 0,1\} ^n} &\to {\{ 0,1\} ^n}\\
x &\to Tag\_COPA(x) = {E_k}({E_k}(x \oplus 5L) \oplus {E_k}(x \oplus 3L) \oplus L) \oplus 7L.
\end{split}
\end{equation}

\begin{figure*}
\centering
  \includegraphics[width=4.5in]{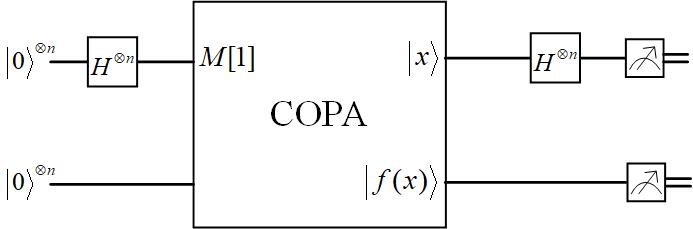}
\caption{The circuit of quantum forgery attack on COPA without associated data ($d=1$).}
\label{fig:4}       
\end{figure*}

The function $f$ can be computed with a single call to the cryptographic oracle, and we can build a quantum circuit for $f$ given a quantum oracle for COPA (as shown in Fig. \ref{fig:4}). Moreover, $f$ satisfies the requirement of Simon's problem with the period $s = 6L$:
\begin{equation}\nonumber
\begin{split}
f(x) &= {E_k}({E_k}(x \oplus 5L) \oplus {E_k}(x \oplus 3L) \oplus L) \oplus 7L,\\
f(x \oplus s) &= {E_k}({E_k}(x \oplus s \oplus 5L) \oplus {E_k}(x \oplus s \oplus 3L) \oplus L) \oplus 7L\\
&={E_k}({E_k}(x \oplus 6L \oplus 5L) \oplus {E_k}(x \oplus 6L \oplus 3L) \oplus L) \oplus 7L\\
&={E_k}({E_k}(x \oplus 3L) \oplus {E_k}(x \oplus 5L) \oplus L) \oplus 7L\\
&=f(x).
\end{split}
\end{equation}
where $6L \oplus 5L = (6 \oplus 5) \cdot L = ({x^2} + x + {x^2} + 1) \cdot L = (x + 1) \cdot L = 3L$ and $6L \oplus 3L = 5L$.

Therefore, the tag of an arbitrary block $M[1]$ is valid for $M[1] \oplus 6L$.

\noindent \textbf{Case 2}. When $d \geqslant 2$, i.e., message $M = M[1]||M[2]|| \cdots ||M[d]$. Then,
\begin{equation}\label{eq04}
T = {E_k}({E_k}(M[1] \oplus M[2] \oplus  \cdots  \oplus M[d] \oplus {2^{d - 1}}{3^2}) \oplus S) \oplus {2^{d - 1}}7L,
\end{equation}
where
\begin{equation}\label{eq05}
S = {E_k}(M[d] \oplus {2^{d - 1}}3L) \oplus  \cdots  \oplus {E_k}(M[2] \oplus 6L) \oplus {E_k}(M[1] \oplus 3L) \oplus L.
\end{equation}

By observing Eqs. (\ref{eq04}) and (\ref{eq05}), we find that if we find the period $s$ which meets $M[1] \oplus s \oplus 3L = M[2] \oplus 6L$, then the tag will not change for different messages $M = M[1]||M[2]|| \cdots ||M[d]$ and $M' = M[1] \oplus s||M[2] \oplus s|| \cdots ||M[d]$, i.e.,
\begin{equation}\nonumber
\begin{split}
S' &= {E_k}(M[d] \oplus {2^{d - 1}}3L) \oplus  \cdots  \oplus {E_k}(M[2] \oplus s \oplus 6L) \oplus {E_k}(M[1] \oplus s \oplus 3L) \\
&\oplus L\\
&={E_k}(M[d] \oplus {2^{d - 1}}3L) \oplus  \cdots  \oplus {E_k}(M[1] \oplus 3L) \oplus {E_k}(M[2] \oplus 6L) \oplus L\\
&=S,\\
T'&= {E_k}({E_k}(M[1] \oplus s \oplus M[2] \oplus s \oplus  \cdots  \oplus M[d] \oplus {2^{d - 1}}{3^2}) \oplus S') \oplus {2^{d - 1}}7L\\
&={E_k}({E_k}(M[1] \oplus M[2] \oplus  \cdots  \oplus M[d] \oplus {2^{d - 1}}{3^2}) \oplus S) \oplus {2^{d - 1}}7L\\
&=T.
\end{split}
\end{equation}

\begin{figure*}
\centering
  \includegraphics[width=4.5in]{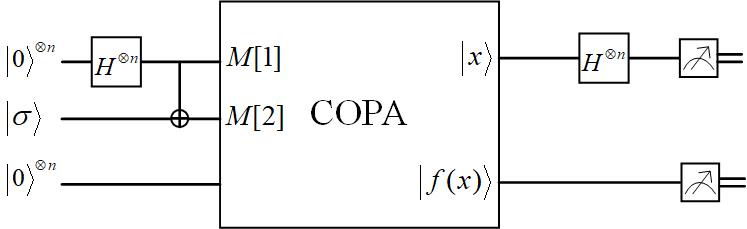}
\caption{The circuit of quantum forgery attack on COPA without associated data ($d=2$).}
\label{fig:5}       
\end{figure*}

Therefore, we only need to intercept the first two message blocks $M[1]||M[2]$ as input to the function $f$, and define the function as below:
\begin{equation}\label{eq06}
\begin{split}
f:{\{ 0,1\} ^n} &\to {\{ 0,1\} ^n}\\
x &\to Tag\_COPA(x||x \oplus \sigma )=\\
&{E_k}({E_k}(\sigma  \oplus 10L)\oplus {E_k}(x \oplus \sigma  \oplus 6L) \oplus {E_k}(x \oplus 3L) \oplus L) \oplus 14L,
\end{split}
\end{equation}
where $10 = {2^{d - 1}} \cdot {3^2} = 2 \cdot {3^2}$, $\sigma  = M[1] \oplus M[2]$ and $\sigma $ can be viewed as an arbitrary constant. It is obvious to see that $f(x) = f(x \oplus s)$ with $s = \sigma  \oplus 5L$:
\begin{equation}\nonumber
\begin{split}
f(x) &= {E_k}({E_k}(\sigma  \oplus 10L) \oplus {E_k}(x \oplus \sigma  \oplus 6L) \oplus {E_k}(x \oplus 3L) \oplus L) \oplus 14L\\
f(x \oplus s) &= {E_k}({E_k}(\sigma  \oplus 10L) \oplus {E_k}(x \oplus s \oplus \sigma  \oplus 6L) \oplus {E_k}(x \oplus s \oplus 3L) \oplus L)\\
 &\oplus 14L\\
 &={E_k}({E_k}(\sigma  \oplus 10L) \oplus {E_k}(x \oplus \sigma  \oplus 5L \oplus \sigma  \oplus 6L) \oplus {E_k}(x \oplus \sigma  \oplus 5L \\
  &\oplus 3L)\oplus L) \oplus 14L\\
  &={E_k}({E_k}(\sigma  \oplus 10L) \oplus {E_k}(x \oplus 3L) \oplus {E_k}(x \oplus \sigma  \oplus 6L) \oplus L) \oplus 14L\\
  &=f(x)
\end{split}
\end{equation}
Therefore, we can apply Simon algorithm on this function $f$ (i.e., the COPA cryptographic oracle), to get period $s = \sigma  \oplus 5L$ (as shown in Fig. \ref{fig:5}). Finally, we query the tag of $M = M[1]||M[2]|| \cdots ||M[d]$, and the same tag is valid for $M' = M[2] \oplus 5L||M[1] \oplus 5L|| \cdots ||M[d]$.

\subsection{Quantum Forgery Attacks on COPA with  {Associated Data}}
\label{sec:3_3}
Due to the existence of associated data, $V = V(A[1]||A[2]|| \cdots ||A[a])$ and $T = {E_k}({E_k}(\Sigma  \oplus {2^{d - 1}}{3^2}L) \oplus {E_k}(M[d] \oplus {2^{d - 1}}3L) \oplus  \cdots  \oplus {E_k}(M[1] \oplus 3L) \oplus V \oplus L) \oplus {2^{d - 1}}7L$. Through the control variable method, we can find that as long as the values of $V$ and $V'$ calculated from two different associated data ($A$ and $A'$, $A \ne A'$) are equal, i.e., $V = V'$, the corresponding tags $T$ and $T'$ with two same constant messages are also equal to each other. So, we can calculate the period $s$ of the function $T$ with the constant message and variable associated data, which is also the period of the function $V$. And by observing the process of processing associated data,  {we find that whether $\left| A \right|$ is a multiple of $n$ will affect the process of calculating} $V$, and $a,d \geqslant 2$. Therefore, We would consider the following three cases: 1) $\left| {A[a]} \right|\% n = 0$, $a = d = 2$; 2) $\left| {A[a]} \right|\% n \ne 0$, $a = d = 2$; 3) $a,b > 2$.

\noindent \textbf{Case 1}: When $\left| {A[a]} \right|\% n = 0$ and $a = d = 2$, i.e., $A = A[1]||A[2]$ and $M = M[1]||M[2]$. Then,
\begin{equation}\label{eq07}
\begin{split}
T = &{E_k}({E_k}(M[1] \oplus M[2] \oplus 10L) \oplus {E_k}(M[2] \oplus 6L) \oplus {E_k}(M[1] \oplus 3L) \\
&\oplus {E_k}({E_k}(A[1] \oplus {3^3}L) \oplus A[2] \oplus 2 \cdot {3^4}L) \oplus L) \oplus 14L.
\end{split}
\end{equation}
In order to prevent the message $M[1]||M[2]$ from affecting the period $s$ of the function $T$ (which is also the period of function $V$ ), we set the value of $M[1]||M[2]$ to an arbitrary constant $m||m$. ${E_k}(M[1] \oplus M[2] \oplus 10L) \oplus {E_k}(M[2] \oplus 6L) \oplus {E_k}(M[1] \oplus 3L)$ would be a constant, which is abbreviated as ${c_{m||m}}$. And Eq. (\ref{eq07}) can be abbreviated as below.
\begin{equation}\label{eq08}
\begin{split}
T = {E_k}({E_k}({E_k}(A[1] \oplus {3^3}L) \oplus A[2] \oplus 2\cdot{3^4}L) \oplus {c_{m||m}} \oplus L) \oplus 14L.
\end{split}
\end{equation}

We firstly fix two arbitrary associated data blocks ${\alpha _0}$ and ${\alpha _1}$ (one of them is $A[1]$ and ${\alpha _0} \ne {\alpha _1}$), and define the following function:
\begin{equation}\label{eq09}
\begin{split}
f:\{ 0,1\}  \times {\{ 0,1\} ^n} &\to {\{ 0,1\} ^n}\\
b,x &\to Tag\_COPA({\alpha _b}||x,m||m)=\\
&{E_k}({E_k}({E_k}({\alpha _b} \oplus {3^3}L) \oplus x \oplus 2\cdot{3^4}L) \oplus {c_{m||m}} \oplus L) \oplus 14L,
\end{split}
\end{equation}
where ${3^3} = {(x + 1)^3} mod({x^{128}} + {x^7} + {x^2} + x + 1)=({x^3} +{x^2}+ x +1) mod({x^{128}} + {x^7} + {x^2} + x + 1)=15$.

\begin{figure*}
\centering
  \includegraphics[width=4.5in]{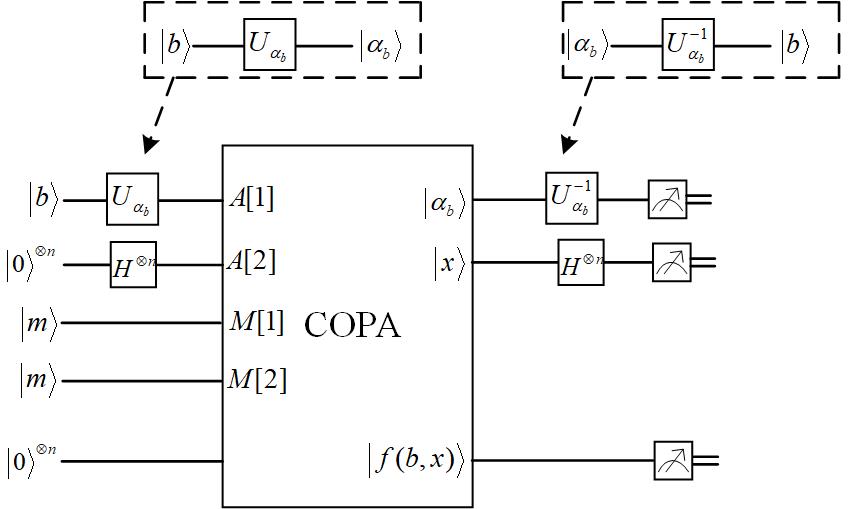}
\caption{The circuit of quantum forgery attack on COPA with associated data ($\left| {A[a]} \right|\% n = 0$, $a=d=2$).}
\label{fig:6}       
\end{figure*}

Then, we apply Simon's algorithm on function $f$ (as shown in Fig. \ref{fig:6}). The function $f$ has the period $s = 1||{E_k}({\alpha _0} \oplus 15L) \oplus {E_k}({\alpha _1} \oplus 15L)$:
\begin{equation}\nonumber
\begin{split}
f(1,x \oplus s) =& {E_k}({E_k}({E_k}({\alpha _1} \oplus 15L) \oplus x \oplus {E_k}({\alpha _0} \oplus 15L) \\
&\oplus {E_k}({\alpha _1} \oplus 15L) \oplus 2\cdot{3^4}L) \oplus {c_{m||m}} \oplus L) \oplus 14L\\
=&{E_k}({E_k}(x \oplus {E_k}({\alpha _0} \oplus 15L) \oplus 2\cdot{3^4}L) \oplus {c_{m||m}} \oplus L) \oplus 14L\\
=&f(0,x).
\end{split}
\end{equation}

When we have got the period of function $V$ (which is also the period of $T$ and $f$ in Eqs. (\ref{eq08}) and (\ref{eq09})) with variable associated data and constant message, we can repeat the attack process in Case 2 of Sect. \ref{sec:3_2} to get another period with different message and no associated data. Finally, the tag of ${\alpha _0}||A[2]$ and $M[1]||M[2]$ is as same as ${\alpha _1}||A[2] \oplus {E_k}({\alpha _0} \oplus 15L) \oplus {E_k}({\alpha _1} \oplus 15L)$ and $M[2] \oplus 5L||M[1] \oplus 5L$'s.

\noindent \textbf{Case 2}: $\left| {A[a]} \right|\% n \ne 0$ and $a = d = 2$, i.e., $A = A[1]||A[2]|{10^*}$ and $M = M[1]||M[2]$. Similar to Case 1 of Sect. \ref{sec:3_3}, we set the value of $M[1]||M[2]$ to an arbitrary constant $m||m$. Then,
\begin{equation}\label{eq10}
\begin{split}
T = {E_k}({E_k}({E_k}(A[1] \oplus {3^3}L) \oplus A[2]|{10^*} \oplus 2\cdot{3^5}L) \oplus {c_{m||m}} \oplus L) \oplus 14L.
\end{split}
\end{equation}
And we define the following function:
\begin{equation}\label{eq11}
\begin{split}
f:\{ 0,1\}  \times {\{ 0,1\} ^n} &\to {\{ 0,1\} ^n}\\
b,x &\to Tag\_COPA({\alpha _b}||x,m||m)=\\
&{E_k}({E_k}({E_k}({\alpha _b} \oplus {3^3}L) \oplus x \oplus 2\cdot{3^5}L) \oplus {c_{m||m}} \oplus L) \oplus 14L.
\end{split}
\end{equation}
Then, we perform Simon's algorithm with this function $f$ (as same as the circuit in Fig. \ref{fig:6}). Its period is as same as the one in Case 1 of Sect. \ref{sec:3_2}, i.e., $s = 1||{E_k}({\alpha _0} \oplus 15L) \oplus {E_k}({\alpha _1} \oplus 15L)$. Finally, the tag of ${\alpha _0}||A[2]|{10^*}$ and $M[1]||M[2]$ is equal to ${\alpha _1}||A[2]|{10^*} \oplus {E_k}({\alpha _0} \oplus 15L) \oplus {E_k}({\alpha _1} \oplus 15L)$ and $M[2] \oplus 5L||M[1] \oplus 5L$'s.

\noindent \textbf{Case 3}: When $a,d > 2$, i.e., $A = A[1]||A[2]|| \cdots ||A[a]$ or $A[1]||A[2]|| \cdots ||A[a]|{10^*}$, and $M = M[1]||M[2]|| \cdots ||M[d]$. Then,
\begin{equation}\label{eq12}
\begin{split}
T = &{E_k}({E_k}(\Sigma  \oplus {2^{d - 1}}{3^2}L) \oplus S \oplus {E_k}({E_k}(A[1] \oplus {3^3}L) \oplus A[2] \oplus  \cdots  \oplus A[a]\\
 &\oplus 2\cdot{3^4}L) \oplus L)\oplus 14L,\\
&or,\\
=&{E_k}({E_k}(\Sigma  \oplus {2^{d - 1}}{3^2}L) \oplus S \oplus {E_k}({E_k}(A[1] \oplus {3^3}L) \oplus A[2] \oplus  \cdots  \oplus A[a]|{10^*}\\
 &\oplus 2\cdot{3^5}L) \oplus L) \oplus 14L.
\end{split}
\end{equation}
Similar to the Case 2 in Sect. \ref{sec:3_2}, we only need to consider the case of $a=3$ to find the period of the function $V$, so $V(A[1]||A[2]||A[3]|| \cdots ||A[a]) = V'(A[1] \oplus s||A[2] \oplus s||A[3]|| \cdots ||A[a])$.

\begin{figure*}
\centering
  \includegraphics[width=4.5in]{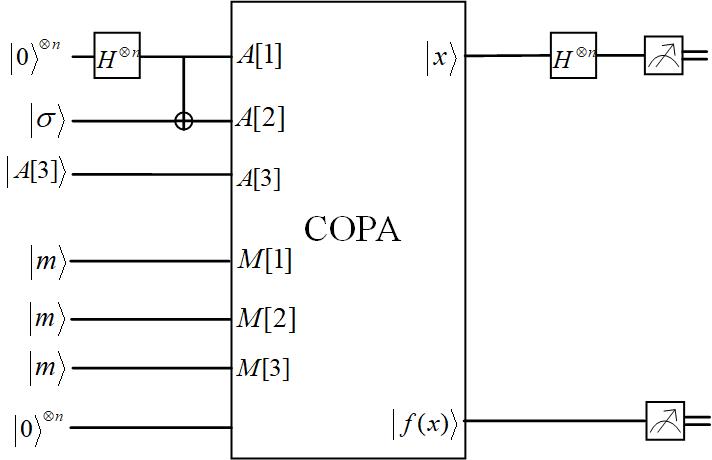}
\caption{The circuit of quantum forgery attack on COPA with associated data ($a,b > 2$).}
\label{fig:7}       
\end{figure*}

We firstly need to calculate the period of $T$ with variable associated data and constant message like Case 1 in Sect. \ref{sec:3_2}, and define the following function:
\begin{equation}\label{eq13}
\begin{split}
f:{\{ 0,1\} ^n} &\to {\{ 0,1\} ^n}\\
x& \to Tag\_COPA(x||x \oplus \sigma ||A[3]{\kern 1pt}or{\kern 1pt} x||x \oplus \sigma ||A[3]|{10^*},m||m||m)   \\
&={E_k}({c_{m||m||m}} \oplus {E_k}({E_k}(x \oplus {3^3}L) \oplus {E_k}(x \oplus \sigma  \oplus 2\cdot{3^3}L) \oplus A[3]\\
&\oplus {\text{ }}{2^2}\cdot{3^4}L) \oplus L) \oplus 14L\\
 &or,\\
&={E_k}({c_{m||m||m}} \oplus {E_k}({E_k}(x \oplus {3^3}L) \oplus {E_k}(x \oplus \sigma  \oplus 2\cdot{3^3}L) \oplus\\
 &A[3]|{10^*} \oplus {2^2}\cdot{3^5}L) \oplus L) \oplus 14L.
\end{split}
\end{equation}
where ${c_{m||m||m}} = {E_k}(\Sigma  \oplus {2^{d - 1}}{3^2}L) \oplus S$ and $\sigma  = A[1] \oplus A[2]$. The circuit of quantum forgery attack is shown in Fig. \ref{fig:7}. We can find that its period is $s = \sigma  \oplus 17L$. Finally, the tag of $A[1]||A[2]|| \cdots ||A[a]{\kern 1pt} {\kern 1pt} {\kern 1pt} or{\kern 1pt} {\kern 1pt} A[a]{\kern 1pt} |{10^*}$ and $M[1]||M[2]|| \cdots ||M[d]$ is identical to $A[2] \oplus 17L||A[1] \oplus 17L|| \cdots ||A[a]{\kern 1pt} {\kern 1pt} or{\kern 1pt} {\kern 1pt} A[a]|{10^*}$ and $M[2] \oplus 5L||M[2] \oplus 5L|| \cdots ||M[d]$'s.

 {From the above quantum forgery attacks on COPA, we found that a quantum superposition query on encrypted Oracle can get all the tags generated by the input plaintext, and then get the orthogonal value of the hidden period through Simon's algorithm. In order to completely compute the hidden period, it is necessary to repeat Simon's algorithm $O(n)$ times. And each time Simon's algorithm is executed, the quantum superposition query needs to be performed again. Therefore, the number of queries for the entire quantum forgery attack is the number of repeated executions of Simon's algorithm. And the success probability of quantum forgery attacks is equivalent to the probability of successfully finding the hidden period through Simon's algorithm.}

\section{Quantum Forgery Attacks on AES-COPA by Simon's Algorithm}
\label{sec:4}
Compared with COPA, AES-COPA has more Nonce (a public message number $N$) as input. And it would participate in the process of processing associated data as part of the associated data, i.e., $X[1]||X[2]|| \cdots ||X[x] = A[1]||A[2]|| \cdots ||A[a]||N$. The rule of processing $X[1]||X[2]|| \cdots ||X[x]$ is as same as the process of processing associated data in COPA. Therefore, the quantum forgery attacks on COPA can also be applied on AES-COPA. Moreover, AES-COPA can accept "fractional" messages $M$, i.e., the length $\left| M \right|$ is not necessarily a positive multiple of the block size $n$. Therefore, we focus on how to implement quantum forgery attack on AES-COPA (v.1 and v.2) with fractional messages in different cases.

\subsection{Quantum Forgery Attacks on AES-COPA v.1}
\label{sec:4_1}
Firstly, assume that we have $d \geqslant 2$ and $1 \leqslant |M[d]| \leqslant n - 1$. Since $XL{S_d}()$ is invertible, $XL{S_d}(M[d]T') \ne XL{S_d}(M{[d]^*}{T'^*})$ for any $M[d] \ne M{[d]^*}$ and $T' \ne {T'^*}$. Therefore, we only to find the period of $T'$ for different messages $M[1]||M[2]|| \cdots ||M[d - 1]$.

For the sake of simplicity, we temporarily set the associated data and Nonce as fixed constants. So, $V$ is a fixed constant, too. And we consider two cases: 1) $d=2$, 2) $d>2$.

\begin{figure*}
\centering
  \includegraphics[width=4.5in]{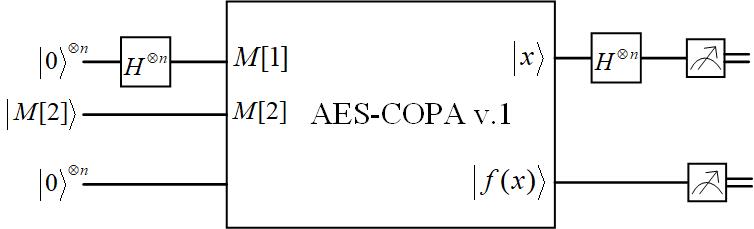}
\caption{The circuit of quantum forgery attack on AES-COPA v.1 ($d=2$, $\left| {M[2]} \right|\% n \ne 0$).}
\label{fig:8}       
\end{figure*}

\noindent \textbf{Case 1}: When $d=2$, $\left| {M[2]} \right|\% n \ne 0$. Then,
\begin{equation}\label{eq14}
  T' = {E_k}({E_k}(M[1] \oplus 5L) \oplus S') \oplus 7L
\end{equation}
where $S' = {E_k}(M[1] \oplus 3L) \oplus V \oplus L$. So,
\begin{equation}\label{eq15}
 T' = {E_k}({E_k}(M[1] \oplus 5L) \oplus {E_k}(M[1] \oplus 3L) \oplus L \oplus V) \oplus 7L
\end{equation}
We can define the following function:
\begin{equation}\label{eq16}
\begin{split}
f:{\{ 0,1\} ^n} &\to {\{ 0,1\} ^n}\\
x &\to Tag\_AES-COPA\_v.1(x) = XL{S_d}({E_k}({E_k}(x \oplus 5L) \oplus {E_k}(x \oplus 3L)\\
 & \oplus L \oplus V) \oplus 7L).
\end{split}
\end{equation}
Through this function, the circuit of our attack on this case is similar to Case 1 in Sect. \ref{sec:3_2} (as shown in Fig. \ref{fig:8}). The period of this function is $s = 6L$. Finally, the tag of $M[1]||M[2]$ is as same as $M[1] \oplus 6L||M[2]$ with same associated data and Nonce. If you want to find two different associated data and Nonce, you can refer to Sect. \ref{sec:3_3}.

\begin{figure*}
\centering
  \includegraphics[width=4.5in]{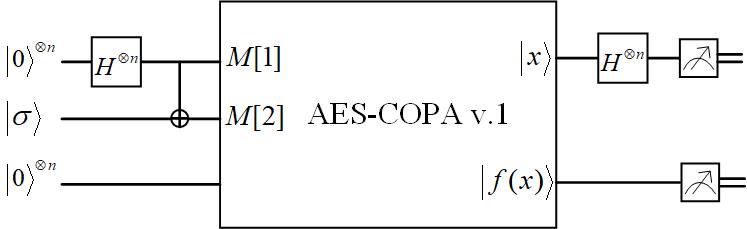}
\caption{The circuit of quantum forgery attack on AES-COPA v.1 ($d>2$, $\left| {M[2]} \right|\% n \ne 0$).}
\label{fig:9}       
\end{figure*}

\noindent \textbf{Case 2}: When $d>2$, $\left| {M[d]} \right|\% n \ne 0$. Then,
\begin{equation}\label{eq17}
  T' = {E_k}({E_k}(M[1] \oplus M[2] \oplus  \cdots  \oplus M[d - 1] \oplus {2^{d - 2}}{3^2}L) \oplus S') \oplus {2^{d - 2}}7L,
\end{equation}
where
\begin{equation}\label{eq18}
 S' = {E_k}(M[d - 1] \oplus {2^{d - 2}}3L) \oplus  \cdots  \oplus {E_k}(M[2] \oplus 6L) \oplus {E_k}(M[1] \oplus 3L) \oplus L \oplus V.
\end{equation}
We can see that this case is similar to Case 2 in Sect. \ref{sec:3_2}. The function can be defined as below:
\begin{equation}\label{eq19}
\begin{split}
f:{\{ 0,1\} ^n} &\to {\{ 0,1\} ^n}\\
x &\to Tag\_AES-COPA\_v.1(x||x \oplus \sigma )=\\
&XL{S_d}({E_k}({E_k}(\sigma  \oplus 10L)\oplus {E_k}(x \oplus \sigma  \oplus 6L) \oplus {E_k}(x \oplus 3L) \oplus L) \oplus 14L)
\end{split}
\end{equation}
where $M[1] \oplus M[2] = \sigma $. Finally, we can apply Simon algorithm on this function $f$ to get period $s = \sigma  \oplus 5L$ (as shown in Fig. \ref{fig:9}). The tag of $M[1]||M[2]|| \cdots ||M[d]$ is valid for $M[2] \oplus 5L||M[1] \oplus 5L|| \cdots ||M[d]$.

In addition to the above two cases, there is a special case, i.e., $d=1$ and $\left| {M[1]} \right|\% n \ne 0$. Due to space limitations, we do not introduce the encryption process of AES-COPA v.1 in this case ( {the entire process can be referred to Ref. \cite{AES-COPAv1}}). And the positions of bits in tag need to be moved in the encryption process, which is not suitable for using the Simon algorithm to implement quantum forgery attacks. Therefore, we do not discuss this case.

\subsection{Quantum Forgery Attacks on AES-COPA v.2}
\label{sec:4_2}
From the entire encryption process of AES-COPA v.2, we can find that the encryption of AES-COPA v.2 for fractional messages is a little different from COPA and AES-COPA v.1. To implement forgery attack, we consider three cases: 1)$d=1$, $\left| {M[1]} \right|\% n \ne 0$; 2)$d=2$, $\left| {M[2]} \right|\% n \ne 0$; 3) $d>2$, $\left| {M[d]} \right|\% n \ne 0$. Assume that the associated data and Nonce are fixed constants.

\noindent \textbf{Case 1}: When $d=1$, $\left| {M[1]} \right|\% n \ne 0$, then,
\begin{equation}\label{eq20}
  T = {E_k}({E_k}(M[1]|{10^*} \oplus {3^2}7L) \oplus {E_k}(M[1]|{10^*} \oplus 3L) \oplus V \oplus L) \oplus 14L
\end{equation}
The function can be defined:
\begin{equation}\label{eq21}
\begin{split}
f:{\{ 0,1\} ^n} &\to {\{ 0,1\} ^n}\\
x &\to Tag\_AES-COPA\_v.2(x) = {E_k}({E_k}(x \oplus 27L) \oplus {E_k}(x \oplus 3L)\\
&\oplus L \oplus V) \oplus 14L,
\end{split}
\end{equation}
where $27 = {3^2}7 = {(x + 1)^2}({x^2} + x + 1) mod ({x^{128}} + {x^7} + {x^2} + x + 1)= {x^4} + {x^3} + x + 1$.

The circuit of our attack in this case is similar to Fig .\ref{fig:4}. Only the cryptographic oracle is the Eq. \ref{eq21} of AES-COPA v.2. The result period is  $s=24L$. So, the tag of $M[1]$ is as same as ${\left\lfloor {M[1]|{{10}^*} \oplus 24L} \right\rfloor _{{{10}^*}}}$, where ${\left\lfloor A \right\rfloor _{{{10}^*}}}$ indicates that the bit string "${10^*}$" at the end of bit string $A$ should be removed. For example, if the last bit in bit string $A$ is "1", then the last bit of $A$ should be removed; if the last two bits in  bit string $A$ are "10", then the last two bits of $A$ should be removed; and so on.

\noindent \textbf{Case 2}:When $d=2$, $\left| {M[2]} \right|\% n \ne 0$, then,
\begin{equation}\label{eq22}
\begin{split}
  T = &{E_k}({E_k}(M[1] \oplus M[2]|{10^*} \oplus 54L) \oplus {E_k}(M[1] \\
  &\oplus 3L) \oplus {E_k}(M[2]|{10^*} \oplus 18L) \oplus V \oplus L) \oplus 28L
\end{split}
\end{equation}
where $54 = 2 \cdot {3^2} \cdot 7$ and $18 = 2 \cdot 3 \cdot 7$.

The function $f$ is defined as below:
\begin{equation}\label{eq23}
\begin{split}
f:{\{ 0,1\} ^n} &\to {\{ 0,1\} ^n}\\
x &\to Tag\_AES-COPA\_v.2(x||x \oplus \sigma) = {E_k}({E_k}(\sigma  \oplus 54L) \\
& \oplus {E_k}(x \oplus 3L) \oplus {E_k}(x \oplus \sigma  \oplus 18L) \oplus V \oplus L) \oplus 28L,
\end{split}
\end{equation}
where $\sigma  = M[1] \oplus M[2]|{10^*}$. Then, we apply Simon's algorithm on this function $f$, whose circuit is similar to Fig. \ref{fig:5}, and get the period $s = \sigma  \oplus 17L$. So, the tag of $M[1]||M[2]$ is equal to ${\left\lfloor {M[2]|{{10}^*} \oplus 17L||M[1] \oplus 17L} \right\rfloor _{{{10}^*}}}$.

\noindent \textbf{Case 3}: When $d>2$, $\left| {M[d]} \right|\% n \ne 0$. Since our attack only requires the first two message blocks $M[1]||M[2]$, whether the last message block is full does not affect implementation of forgery attacks. The process of our quantum forgery attack is as same as Case 2 in Sect. \ref{sec:4_2}.

 {Similar to the quantum forgery attacks on COPA, under the Q2 model, the number of quantum superposition queries is the number of times that Simon's algorithm is executed, and the success rate of our attack is also the success rate of the hidden period calculated.}

\section{Quantum Forgery Attacks on Marble by Simon's Algorithm}
\label{sec:5}
By observing the entire overview of Marble v1.2, we find that the attack strategy in Sect. \ref{sec:3} can not apply on Marble directly. But inspired by Lu's attack strategy \cite{Lu17}, we can use Simon algorithm to recover the value of $L$ firstly, and then calculate the tag of new message by $L$ to achieve the purpose of forgery attack.

\begin{figure*}
\centering
  \includegraphics[width=4.5in]{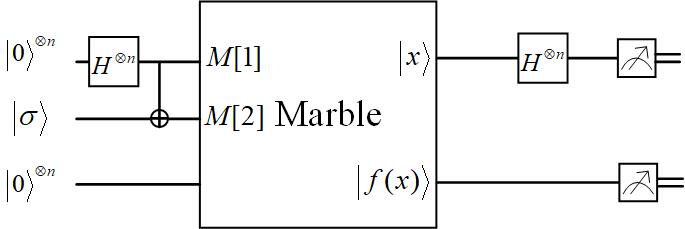}
\caption{The circuit of quantum forgery attack on Marble.}
\label{fig:10}       
\end{figure*}

To recover the value of $L$, we can apply Simon algorithm on the process of calculating ${S_1}$ (${S_1}$ in Marble v1.2):
\begin{equation}\label{eq24}
\begin{split}
{S_1} =& Cons{t_1} \oplus {E_1}(Cons{t_0}) \oplus {E_1}(A[1] \oplus 5L) \oplus {E_1}(A[2] \oplus 10L) \oplus  \cdots  \\
&\oplus {E_1}(A[a] \oplus {2^{a - 1}}{3^3}L) \oplus {E_1}(M[1] \oplus 2L) \oplus {E_1}(M[2] \oplus 4L) \oplus  \cdots \\
 &\oplus {E_1}(M[d] \oplus {2^d}L) \oplus {E_1}(\Sigma  \oplus \tau  \oplus {2^d} \cdot 7L),
\end{split}
\end{equation}
For easy computing, we do not consider the associated data and choose an arbitrary constant $\sigma $, and then define the following function:
\begin{equation}\label{eq25}
\begin{split}
f:{\{ 0,1\} ^n} &\to {\{ 0,1\} ^n}\\
x &\to {S_1}\_Marble(A = 0,M = x||x \oplus \sigma )=\\
&Cons{t_1} \oplus {E_1}(Cons{t_0}) \oplus {E_1}(x \oplus 2L) \oplus {E_1}(x \oplus \sigma  \oplus 4L) \oplus {E_1}(\sigma  \oplus 14L)
\end{split}
\end{equation}
where $Cons{t_0}$ and $Cons{t_1}$ are constants. Finally, we can use Simon algorithm to recover the function $f$'s period $s = \sigma  \oplus 6L$ (as shown in Fig. \ref{fig:10}). The value of $L$ can be obtained by $s$.

When we get the value of $L$, we can query the Marble v1.2 encryption oracle with $d + 1$ block message $\tilde M = M[1]||M[2]|| \cdots ||M[d]|| \oplus _{i = 1}^d{M_i} \oplus {2^{d + 1}}L \oplus {2^d} \cdot 7L$, and obtain its ciphertext $\tilde C = (C[1],C[2], \cdots ,C[d],\tilde C[d + 1])$. Then, the ciphertext for $M = M[1]||M[2]|| \cdots ||M[d]$ is $C = (C[1],C[2], \cdots ,C[d])$, and the tag for $M$ is $\tilde C[d + 1] \oplus {2^d} \cdot 3 \cdot L \oplus {2^{d - 1}} \cdot 3 \cdot 7 \cdot L$. Since the associated data is not considered, the ciphertext and tag are also applicable to the Marble v1.1. The specific process of generating forged tag can refer to Ref. \cite{Lu17}.

On the other hand, if the associated data is considered, the forgery attack can refer to Fuhr \textit{et al.}'s attack strategy \cite{Fuhr15}. For any associated data $A$ and message $M$, the adversary computes the masked value $B$ of a chunk of 8 identical blocks of associated data after $A$ and queries the encryption oracle on $\left( {A||B,M} \right)$. The answer $\left( {C,T} \right)$ to that query is also valid ciphertext and tag for $\left( {A,M} \right)$. The attack for Marble v1.0 is also applicable to other versions of Marble.

 {Different from the quantum forgery attacks on COPA and AES-COPA, the quantum forgery attacks here are mainly to obtain the secret parameter $L$ through quantum superposition query and Simon's algorithm. But the attack process is similar. So the number of queries and the success rate are the same.}

\section{Efficiency Comparison}
\label{sec:6}
In this section, we will compare our quantum forgery attack with other forgery attacks (Ref.\cite{Nandi15,Lu17,Fuhr15}) in terms of attack efficiency, which mainly consists of two aspects: the number of queries and the success probability of forgery attacks. The comparison result is shown in Table \ref{Tab:3}. In Ref. \cite{Lu17,Fuhr15}, they use birthday attack to find the collisions of the tag, which can recover the secret parameter $L$. Then, they can compute the forgery tag by querying the oracle once for specified message. We can see that the number of queries mainly concentrate in recovering the secret parameter $L$. So, the number of queries is close to the birthday-bound ${2^{n/2}}$. And Ref. \cite{Nandi15} uses pigeonhole principle on the case of processing fractional messages to implement forgery attack, which only needs ${2^{n/3}}$ queries. Their success probability is a trade-off with the number of queries.

In our quantum forgery attack, we rely on the Simon's algorithm to query collisions. Therefore, the number of queries in our attack is roughly equivalent to the number of repeating Simon's algorithm. And the success probability of our attack is equal to the success probability of Simon's algorithm. Due to the characteristics of the queries in superposition, the number of queries can be reduced from exponential time to polynomial time. Our attack only needs $cn$ queries, where $c$ is a constant. Besides, Shi \textit{et al.} \cite{Shi19} proved that Simon's algorithm returns $s$ with $cn$ queries, with probability at least $1 - {2^n} \times {(0.6454)^{cn}}$. If we choose $c=4$ and $n=128$, the success probability is very close to 1, which is much greater than these classic forgery attack.

\begin{table}[ht]
  \caption{Comparison between classic forgery attack and quantum forgery attack}
  \label{Tab:3}
  \centering
  \begin{tabular}{p{1cm}p{1.5cm}p{3cm}p{1cm}p{1cm}p{2cm}}
  \toprule
  & & Ref. \cite{Lu17}& Ref. \cite{Nandi15}& Ref. \cite{Fuhr15} & Our attack\\
  \midrule
  \multirow{2}{1cm}{COPA}& Queries & ${2^\sigma } + {2^\varphi }$ $\left( {1 \leqslant \sigma ,\varphi  < \frac{n}{2}} \right)$ & ${2^{n/3}}$ & \# & $cn$ \\
  &Success Prob. & $1 - {e^{ - {2^{\sigma  + \varphi  - n}}}}$ &25\%  & \# &$1 - {2^n} \times{(0.6454)^{cn}}$ \\
  \hline
  \multirow{2}{1cm}{AES-COPA (v.1/2)}& Queries &${2^{63}}$ &\# &\# &$cn$\\
  &Success Prob. &6\% &\# &\# &$1 - {2^n} \times{(0.6454)^{cn}}$\\
  \hline
  \multirow{2}{1cm}{Marble (v1.0/1/2)}& Queries & ${2^{65}}$ &\# & ${2^{65}}$& $cn$\\
  &Success Prob. &32\% &\# & 32\%& $1 - {2^n} \times{(0.6454)^{cn}}$\\
  \bottomrule
  \end{tabular}
\end{table}

\section{Discussion and Conclusion}
\label{sec:7}
In this paper, we have presented quantum forgery attacks on COPA, AES-COPA and Marble authenticated encryption algorithms. Due to quantum superposition query and Simon's algorithm, our quantum forgery attack on COPA, AES-COPA and Marble are more efficient than classic forgery attacks, i.e., it only needs $cn$ queries and its success probability is very close to 1.

However, the premise for our quantum forgery attack to be effective is in the quantum setting. If the attacker only queries classically, we may not reduce the queries to only $O(n)$ times. But we can use Grover algorithm offline to find collisions, which can reduce the queries from $O({2^{n/2}})$ to $O({2^{n/3}})$. So, using quantum algorithms offline to improve the efficiency of breaking symmetric cipher will be one of our future research directions.

\begin{acknowledgements}
This work was supported by the National Natural Science Foundation of China under Grant 61672290 and Grand 61802002, the Natural Science Foundation of Jiangsu Province under Grant BK20171458, the Graduate Research and Innovation Projects of Jiangsu Province (KYCX20\_0978), and the Natural Science Foundation of the Jiangsu Higher Education Institutions of China (19KJB520028), and in part by the Priority Academic Program Development of Jiangsu Higher Education Institutions (PAPD).
\end{acknowledgements}

%
%



\end{document}